\documentclass[12pt]{article}
\usepackage{hyperref} 

\hypersetup{
    colorlinks,
    linkcolor={red!50!black},
    citecolor={green!50!black},
    urlcolor={blue!80!black}
}
\usepackage{comment}
\usepackage{longtable} 
\usepackage{xcolor} 
\usepackage{amsmath, amssymb, amsthm}
\usepackage{mathrsfs} 
\usepackage{graphicx,psfrag,epsf}
\usepackage{enumerate}
\usepackage[utf8]{inputenc}
\usepackage{graphics}
\usepackage{subcaption} 

\usepackage{geometry}

\usepackage{newunicodechar}
\newunicodechar{ℝ}{\mathbb{R}}
\usepackage{algorithm}
\usepackage{algorithmic}
\usepackage{array}
\usepackage{multirow} 
\usepackage{cleveref} 
\crefformat{equation}{(#2#1#3)}

\usepackage{algorithm}
\usepackage{algorithmic}

\theoremstyle{plain} 

\usepackage{graphicx}
\usepackage{subcaption}
\usepackage[export]{adjustbox}

\newcommand{\blind}{0}

\usepackage{caption}
\begin{document}

\def\spacingset#1{\renewcommand{\baselinestretch}%
{#1}\small\normalsize} \spacingset{1}


\if0\blind
{
  \title{ Evaluation of Minimal Residual Disease as a Surrogate for Progression-Free Survival
in Hematology Oncology Trials: A Meta-Analytic Review}
\author{
 \textbf{Jane She} \\
  Department of Biostatistics, University of North Carolina, \\
  135 Dauer Drive, Chapel Hill, North Carolina 27599, U.S.A. \\
  jane.she@unc.edu \\
   \and
   \textbf{Xiaofei Chen} \\
   Oncology Biometrics, Astrazeneca \\
   xiaofei.chen2@astrazeneca.com \\
   \and
   \textbf{Malini Iyengar} \\
   Oncology Biometrics, Astrazeneca \\
   malini.iyengar@astrazeneca.com
   \and
   \textbf{Judy Li} \\
   Oncology Biometrics, Astrazeneca \\
   judy.li1@astrazeneca.com
}
  \maketitle
} \fi

\if1\blind
{
  \bigskip
  \bigskip
  \bigskip
  \begin{center}
    {\small Title}
\end{center}
  \medskip
} \fi

\bigskip
\begin{abstract}

Traditional health authority approval for oncology drugs is based on a clinical benefit endpoint, or a valid surrogate. In 1992 FDA created the Accelerated Approval pathway to allow for earlier approval of therapies in serious conditions with an unmet medical need. This is accomplished typically by granting an accelerated approval based on surrogate endpoint, that can be measured earlier than a traditional approval endpoint. Minimal residual disease (MRD) is a sensitive measure of residual cancer cells in hematology oncology after treatment, and is increasingly considered as a secondary or exploratory endpoint due to its prognostic potential for traditional clinical trial endpoints such as progression-free survival (PFS) and overall survival (OS). This work aims to evaluate MRD’s surrogacy potential across several hematology cancer indications (FL), and others while keeping the focus on follicular lymphoma, using data from published studies. We examine individual-level and trial-level correlations extracted from previously published studies to elucidate the potential role of MRD in accelerating the drug approval process in hematology oncology trials \cite{buyse2000validation}.

\end{abstract}

\vspace{5mm}

\noindent%
{\it Some keywords:} Minimal residual disease, surrogacy, progression-free survival, hematology oncology, meta-analysis

\spacingset{1.45}
\section{Introduction}
\label{sec:intro}

Minimal residual disease (MRD) is a measurement of cancer cells remaining in the blood after treatment that can be detected at a minuscule level. Presence of  residual cancer cells after treatment can potentially result in relapse but are conventionally undetectable without recent technological advances. Various assays with specific thresholds of sensitivity have been developed to measure MRD including next generation genetic sequencing, flow cytometry, and real-time quantitative reverse transcriptase polymerase chain reaction (RQ-PCR). Current standards can detect presence of cancer cells at levels from $1:10^4$ to $1:10^6$ nucleated cells, whereas conventional cytomorphology using peripheral blood smears and bone marrow aspirations may only reach sensitivity levels of $1:20$ (Leukemia \& Lymphoma Society). 

Due to its potential as a prognostic marker as well as its earlier use capabilities compared to traditional clinical endpoints such as progression-free survival (PFS) and overall survival (OS), MRD is increasingly being included across various hematology oncology indications as an exploratory endpoint for response assessment, which raises questions about its ability to act as a surrogate endpoint for PFS or OS, since it has already been granted as an intermediate endpoint for multiple myeloma (MM) \cite{FDA_ODAC2024}. In follicular lymphoma (FL), B-cell acute lymphoblastic leukemia (B-ALL), mantle cell lymphoma (MCL), and chronic lymphocytic leukemia (CLL), MRD negativity has been shown in several trials to be an excellent predictor for patient survival outcomes \newline
\cite{zhou2022clinical, pott2024minimal, kaplan2020bortezomib, bottcher2012minimal}.

Recently, the U.S. Food \& Drug Administration’s (FDA) Oncologic Drugs Advisory Committee (ODAC) approved MRD as an early endpoint for multiple myeloma (MM) to support accelerated approvals for new therapies \cite{FDA_ODAC2024}. Before this announcement, the typical endpoints used in MM trials were based on PFS and OS. These endpoints, used for regular FDA approval, typically take many years of patient follow-up to collect enough data for measurement and use, which prolongs the drug development process and delays patient access to novel treatments.

With MRD being approved as an early endpoint for AA in MM, the drug approval process is expedited, as MRD can be measured within $1$ year, and was demonstrated via independent meta-analyses to have strong individual-level associations for both PFS and OS, and weak to moderate trial-level associations. This means that though MRD is not a validated surrogate endpoint, in that it cannot fully replace PFS and OS, it is a strong prognostic factor which can be measured sooner to enable patients to access new treatments faster while sponsors continue confirmatory trials.

\begin{figure}[H]
    \centering
    \includegraphics[width=\linewidth]{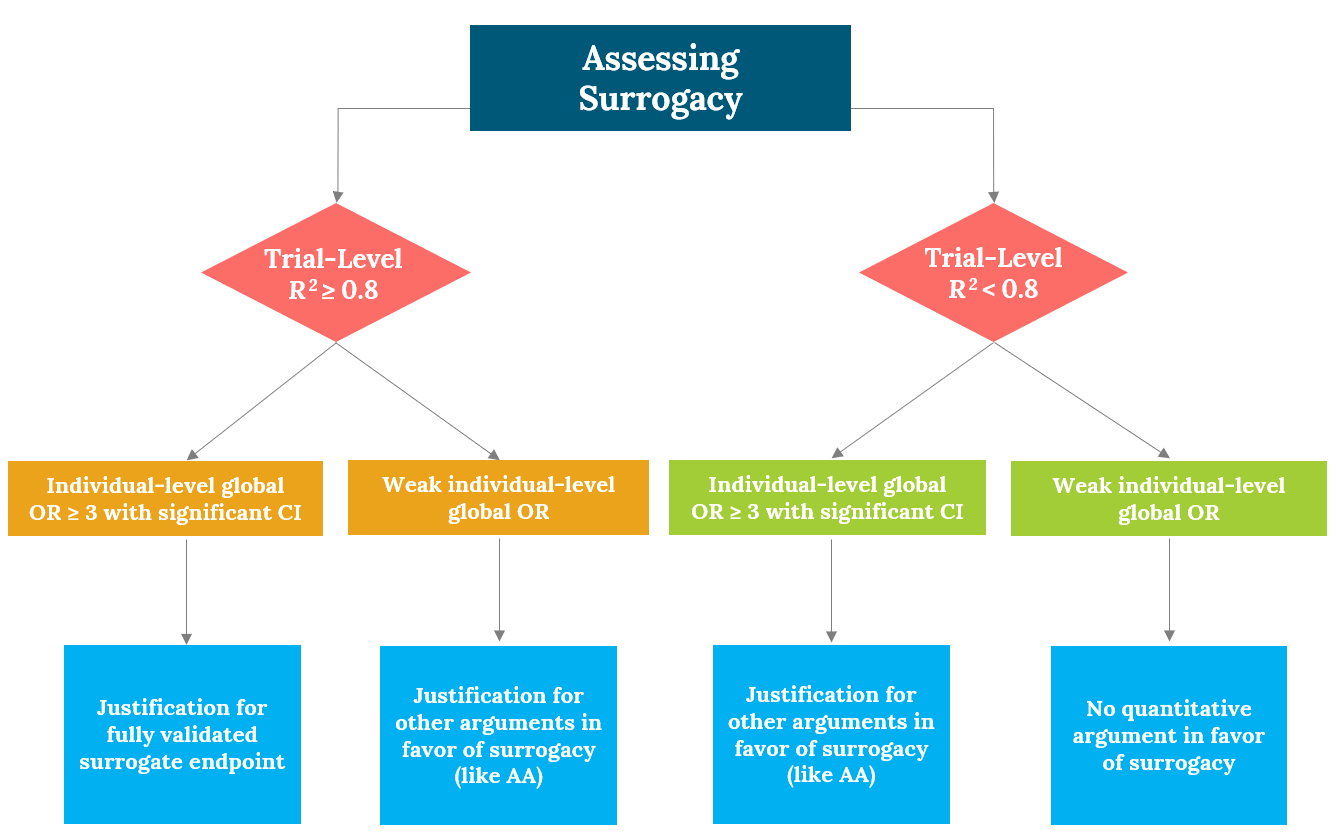}
    \caption{Flowchart for Surrrogacy Assessment}
    \label{fig:surr}
\end{figure}

There is no current FDA guided standard threshold to define a ``strong" trial-level correlation, with previous meta-analyses citing predefined $R^2_{Copula}$ \& $R^2_{WLS}$ values of $\geq 0.8$ as desirable with a lower bound of the $95\%$ confidence interval $> 0.6$, and neither $R^2$ estimate $< 0.7$ \cite{dixon2022end, shi2017thirty, shi2018progression, yin2022reevaluating}. This criteria is the same that the International Independent Team for Endpoint Approval (i$^2$TEAMM) used in the analysis of MRD as a surrogate for MM 
\newline \cite{FDA_ODAC2024}, and is what this paper will proceed with as well, in absense of regulatory guidance on these thresholds. This can be visualized in Figure \ref{fig:surr}. \\ For more details about the process of defining surrogate endpoints and intuition behind the process in the figure, please refer to the supplemental material.

\cite{galimberti2018validation} explored validation of MRD as a surrogate endpoint for EFS in pediatric ALL, however, there is a need for exploration in adult ALL as well. \cite{narkhede2018surrogate} mention the possibility of using MRD in place of PFS in FL through a clinical lens, but do not go through the process of surrogacy validation. \cite{dimier2018model} examine surrogacy of MRD at the end of induction after treatment with chemoimmunotherapy to act as a surrogate for PFS using $3$ trials, and extracting subgroups from the trials to act as data points. However, the paper does not include more recent publications for two-arm trials in the 1L setting that have been included in our analysis. Additionally, the aforementioned analysis only assesses trial-level correlation, and not individual-level correlation.

Similarly, \cite{simon2024endpoint} recently evaluated surrogacy of MRD for PFS in first-line CLL using both trial-level and individual-level correlation, with the conclusion that there was a strong patient-level correlation between PFS and MRD, but uncertain trial-level correlation. However, our analysis differs in the use of the plackett copula for model fitting as opposed the joint-frailty copula, where the plackett copula model has an added ease of interpretation with a global odds ratio which quantifies how the risk of one event is related to the risk of another. This means that the correlations we obtain are directly interpretable, adding an extra layer of credibility.

Section \ref{sec:meth} includes details about the literature search, literature selection, and data extraction process as well as details about the statistical methodology employed. Next, section \ref{sec:results} presents results for each indication. The overall conclusions are then discussed in section \ref{sec:disc}.

\section{Materials and Methods}
\label{sec:meth}

\subsection{Literature Search \& Review}

Literature search for each indication was conducted by the Information Practice (IP) team at Astrazeneca PLC, searching for studies reporting MRD based endpoints   along with survival outcomes which were published in \href{https://www.embase.com/landing?status=grey}{\textit{embase}}
, \href{https://www.nlm.nih.gov/medline/medline_home.html}{\textit{medline}}, \href{https://clinicaltrials.gov/study/NCT06268886}{\textit{clinicaltrials.gov}}, or \href{https://www.citeline.com/en/products-services/clinical/trialtrove?utm_source=google&utm_medium=cpc&utm_term=trialtrove|&utm_campaign=SO-Clinical-AMER&gad_source=1&gclid=EAIaIQobChMI4OO3xrPGjAMV6mtHAR3OGTddEAAYASAAEgKtrPD_BwE%20is%20this%20the%20one?%20If%20so%20we%20can%20cr}{\textit{trialtrove}}. The searches were conducted between February and August 2024. 

Studies that did not present results of therapeutic evaluation were excluded, as were studies that did not report survival curves which were in some way, stratified by MRD status, as the information would not be complete for extraction. For this meta-analysis to establish surrogacy, only two-arm trials were considered for use to allow for estimation of treatment effects. However, if there were not enough two-arm trials in an indication, one-arm trials were considered in forming individual-level correlations between the true endpoint and the surrogate endpoint. This will be further described in section \ref{sec:onearm}.

For eligible studies, the following information was collected: year of publication, population, disease setting (1L, R/R etc.), eligibility criteria, interventions evaluated, survival outcomes of interest (PFS, OS, EFS, etc.), MRD measurement time frame, MRD measurement sensitivity.

\subsection{Data Extraction}

Survival data was extracted using the ``ScanIt" free software, which allows for extraction of survival curves via a pixel-based method. This was then converted into individual patient data (IPD) using \textsf{R} version 4.4.0 with package ``IPDfromKM", which implements a modified-iKM algorithm for IPD reconstruction through a two-stage process (more details see \cite{liu2021ipdfromkm}). The first stage involving extracting data coordinates, is accomplished via the ``ScanIt" software; in the second stage, these coordinates are preprocessed into IPD data. 

The original survival curves from the initial data extraction can be reconstructed visually once IPD is constructed, as a form of verification. The hazard ratios of PFS between the two treatment arms were calculated by fitting a Cox proportional hazard model with the treatment as the explanatory variable.

Additionally, odds ratios of MRD negativity were calculated. If the sample size from summing all cells of an odds ratio calculation was different from the sample size resulting from the IPD, the smaller sample size was used for weighting to be more conservative. Details about the quality control process are discussed in the supplement.

\subsection{Evaluation of Surrogacy}

Overall surrogacy was evaluated by considering both the individual-level and trial-level associations through meta-analysis using the framework proposed by \cite{buyse2000validation}.Based on the current evidence, the goal is to assess whether MRD has the potential to be a validated surrogate endpoint in hematology oncology trials, and may be used in traditional FDA drug approval, or as an intermediate endpoint to be used in accelerated approval, or if it has no evidence of potential surrogacy.  \cite{haslam2019systematic} conducts a systematic review of some trial-level meta-analyses for surrogate endpoints of OS in oncology and \cite{belin2020progression} systematically examines the methodology used in assessing PFS as a surrogate for OS.

The methods employed in this analysis included three models to assess surrogacy at the trial-level. Two models employed weighted least squares (WLS) modeling and used the log hazard ratio of PFS as the response, and the log odds ratio of MRD negativity as the predictor.  As only two arm trials were considered, the hazard ratio of PFS events compares the hazards of failure between the experimental arm and the control/active comparator. The odds ratio of MRD negativity uses the same treatment arm comparisons for consistency. These were weighted by the inverse variance of the log odds ratio for the first model and sample size for the second model, as a sensitivity analysis to adjust for trial size, and result in RWLS2 values for interpretation along with a $95\%$ confidence interval. In statistics, the $R^2$ value is frequently referred to as the coefficient of determination, which takes on values between $0$ and $1$ and denotes the proportion of variation in the outcome that can be explained by the variation in the predictor.

The third model at the \underline{trial-level} is fit using a bivariate Plackett copula model, as the first two approaches do not account for estimation error of the trial-specific effects \cite{belin2020progression}. Additionally, this model accounts for patient-level correlation between the true endpoint such as PFS and MRD. The first stage of the model estimates trial-specific effects on the surrogate and the true endpoint using a fixed-effects model. As MRD is a binary endpoint, this is equivalent to using logistic regression with a Weibull model, where the logistic regression is used to estimate the treatment on MRD and the Weibull model is used to estimate the treatment on the hazard rate of PFS \cite{burzykowski2001validation}. The goal of the first stage is to estimate the effect sizes and their standard errors for each trial using a likelihood-based approach. The second stage then uses the first-stage parameter estimates in a trial-level model to estimate association between treatment effects and assess the quality of the surrogate by squaring the association to calculate a $R^2_{Copula}$ value.  This value assesses the correlation between the treatment effect on the surrogate endpoint and the true endpoint using a copula-model. 

At the \underline{individual-level}, the same bivariate Plackett copula model is used which results in a ``global odds ratio", an odds ratio of two odds. The numerator consists of the odds of having a PFS event at a certain time for patients who are MRD negative; the denominator consists of the odds of having a PFS event at a certain time for patients who are MRD positive. For example, a value higher than $1$ indicates patients who are MRD negative have longer PFS outcomes, and that the odds of being alive and without progression for MRD negative patients is higher than the odds of patients who are MRD positive at the same timepoint.

For the analysis, the package ``Surrogate" in \textsf{R} was used to fit the bivariate plackett copula model, and the package ``Metafor" was used to conduct the WLS regressions, as \cite{thompson1999explaining} recommend using methods with additive error models rather than multiplicative error models \cite{viechtbauer2010conducting}. To calculate confidence intervals for the two weighted least squares approaches, bootstrapping with $10,000$ replicates was performed to sample points with replacement and fitting a linear model. The $95\%$ confidence intervals were calculated by extracting the $2.5$ and $97.5$ quantiles of the bootstrapped $R^2$ values. However, this approach is limited by the small sample size; so confidence intervals have a significant amount of variability between analyses, resulting in either near-perfect fits, or values essentially showing $[0, 1]$ depending on the selection of points.

A limitation to the use of the package ``Surrogate" fitting the bivariate plackett copula model, is that when the estimated correlation parameter in the second-stage of the model has confidence intervals spanning both negative and positive values, the output values of the $R^2_{Copula}$ with its $95\%$ confidence intervals may be misleading., resulting in a lower bound larger than the upper bound. Take, for example, a correlation point estimate of $0.1$, with confidence interval of $[-0.8, 0.6]$. This then becomes $R^2_{Copula} = 0.1^2 = 0.01$ with a CI of $[0.64, 0.36]$. The lower bound exceeds the upper bound, so the reported CI should actually be reported as $[0, 0.64]$. The same should be said for any correlation CIs that span both negative and positive values, where the lower bound of the squared values should be $0$, as $0$ is included in the correlation CI, and the upper bound should be the larger of the two values.

\subsection{One-arm trials}
\label{sec:onearm}

If there were not enough literature available for a two-arm surrogacy analysis as planned, single arm papers were considered to form an association metric between the surrogate endpoint(MRD) and the true endpoint (PFS/OS). This is distinct from the estimation of a trial-level correlation, as this association excludes treatment effects due to the lack of a control arm. More details are discussed in the supplemental section.

\section{Results}
\label{sec:results}

A table of the comprehensive literature review performed is summarized in Table \ref{tab:lit_rev}, with the number of evaluated papers encompassing all settings for papers involving MRD evaluation and survival outcomes. The setting presented refers to the setting for the number of usable trials, which are trials with MRD stratified survival curves with the survival endpoint of interest.

\begin{table}[h]
    \centering
    \caption{MRD outcomes landscape summary for relevant survival endpoints}
    \captionsetup{justification=centering}
    \scalebox{0.9}{
    \begin{tabular}{ccccc}
        \hline
        Indication & Setting & Two-Arm Trials (N) & One-Arm Trials (N) & Papers Evaluated (N) \\
        \hline\hline
        FL & 1L & 4 (PFS) & -- & 19 \\
        ALL &  R/R & 0 & 3 (OS) & 17 \\
        MCL & 1L & 0 & 6 (3 OS, 3 PFS) & 25\\
        CLL & 1L & 4 (PFS) & -- & 41\\
        \hline
    \end{tabular}
    }

    \caption*{\footnotesize FL = Follicular Lymphoma ; ALL = Acute Lymphocytic Leukemia;  MCL = Mantle Cell Lymphoma, CLL = Chronic Lymphocytic Leukemia}
    \label{tab:lit_rev}
\end{table}

\vspace{5mm}

\underline{Follicular Lymphoma (FL)}\\

Through the literature search and data extraction criteria for FL, five data points were extracted from the following papers: \cite{pott2024minimal, delfau2020lenalidomide, bishton2020uk, pott2020mrd, luminari2021response}. All trials were conducted in a phase III setting, with four trials in the 1L setting and one trial in the R/R setting. As the patient population may be different for patients who are previously untreated as compared to relapsed, we excluded the R/R trial, which is known as the ``GADOLIN" trial, leaving four data points for use. Characteristics of the extracted trials can be seen in Table \ref{tab:char_FL}. 

The analysis was conducted with all four points included as well as with a sensitivity analysis excluding the FOLL12 trial \cite{luminari2021response}. This is because the FOLL12 trial resulted in an experimental arm, which treated patients with response-oriented post-induction therapy, presenting worse outcomes than the standard arm, which uses rituximab maintenance. The PFS of the experimental arm was significantly poorer than the PFS of the standard arm. The inclusion of this trial introduces a large level of heterogeneity, as the sample size is large and thus, the trial carries large weight on the results.

\begin{table}[h]
    \centering
    \caption{FL Trial Characteristics and Assessment Results for FL}
    \begin{subtable}{\linewidth}
        \centering
        \scalebox{0.8}{
        \begin{tabular}{lccccc}
            \hline
            Trial Name & Setting/Phase & Sample Size & \multicolumn{1}{p{4cm}}{\centering MRD Measurement \\ Time (months) } &\multicolumn{1}{p{5cm}}{\centering MRD Measurement \\ Method/Sensitivity } & Treatments \\
            \hline\hline
            GALLIUM &1L Phase 3 & 693 & 9-12 &PCR $10^{-4}$ & G-chemo vs R-chemo \\
            RELEVANCE &1L Phase 3 &  207 & 5.52 &PCR $10^{-4}$& $R^2$ vs R-chemo \\
            UK NCRI &1L Phase 3 &  87 & 7.36 &PCR $10^{-4}$& CMD vs FMD  \\
            FOLL12 &1L Phase 3 &  345 & 4.14-5.52 &PCR $10^{-4}$& RM vs response-adapted \\
            \hline
        \end{tabular}
        }
        \caption{FL trial characteristics}
        \label{tab:char_FL}
    \end{subtable}

    \vspace{0.5cm} 

    \begin{subtable}{\linewidth}
        \centering
        \scalebox{0.8}{
        \begin{tabular}{lcc}
            \hline
            Correlation Metrics & All Trials & Excluding FOLL12 \\
            \hline\hline
            \textbf{Trial-Level} & & \\
            Sample Size & 1332 & 987 \\
            $R^2_{WLS}$ inverse variance ($95\%$ CI) & 0 [0, 1.000] & 0.98 [0, 1.000]\\
            $R^2_{WLS}$ sample size ($95\%$ CI) & 0.247 [0, 1.000]  & 0.962 [0, 1.000]\\
            $R^2_{Copula}$ ($95\%$ CI) &  0.244 [0, 0.826] & 0.974 [0.778, 0.997]\\
            \hline\textbf{Individual-Level} & & \\
            Sample Size & 1332 & 987 \\
            Bivariate Plackett Copula Global OR ($95\%$ CI) &  2.466 [1.672, 3.259] & 2.709 [1.715, 3.703]\\
            \hline
        \end{tabular}
        }
        \caption{Trial/individual-level assessment results for FL}
        \label{tab:results_FL}
    \end{subtable}
\end{table}

All trials for FL had a MRD sensitivity threshold of minimum $10^{-4}$ and were measured using PCR, with MRD status being measured at the end of induction (EOI) for three of the 1L studies, and at the end of treatment in the other 1L study. However, EOI may be dependent on the treatment within a study, thus, there were ranges reported in terms of months in Table \ref{tab:char_FL}. Therefore, the time frame in which MRD was measured ranges from approximately $4.14$ months to 12 months. Additionally, the median follow-up time across all patients with all trials included is $45.4$ months, or approximately $3.78$ years. In the sensitivity analysis, with the FOLL12 trial removed, the median follow-up time across all patients is $47.5$ months, or approximately $3.96$ years.

\begin{figure}[H]
    \centering
        \includegraphics[width=6in]{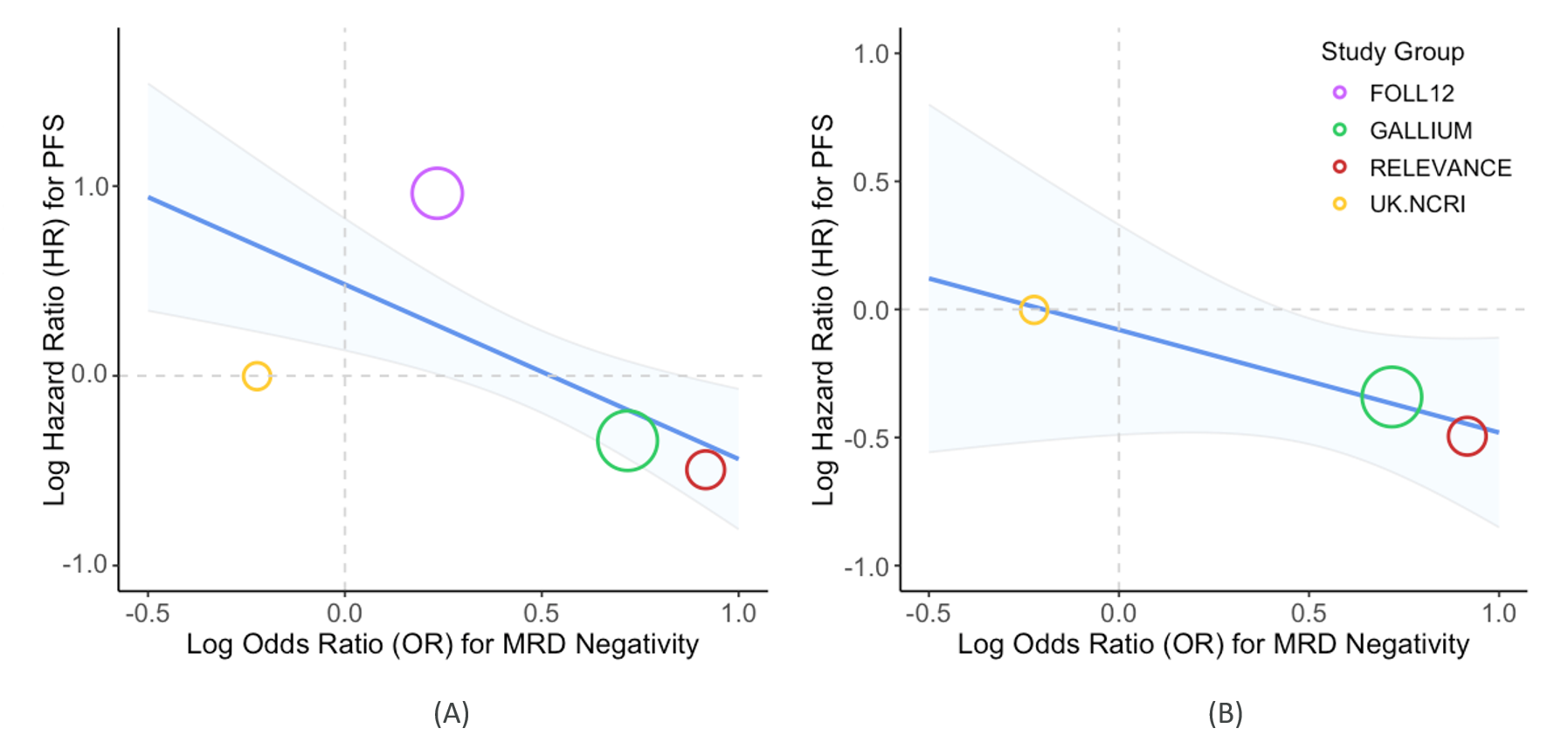} 
    \caption{\textbf{Weighted by inverse variance of log OR}: Correlation between treatment effect on MRD negativity and treatment effect on PFS. (A) all points: WLS for FL with all points included; (B) sensitivity analysis: WLS for FL with FOLL12 excluded}
    \label{fig:FL_wls_var}
\end{figure}

\begin{figure}[H]
    \centering
        \includegraphics[width=5.8in]{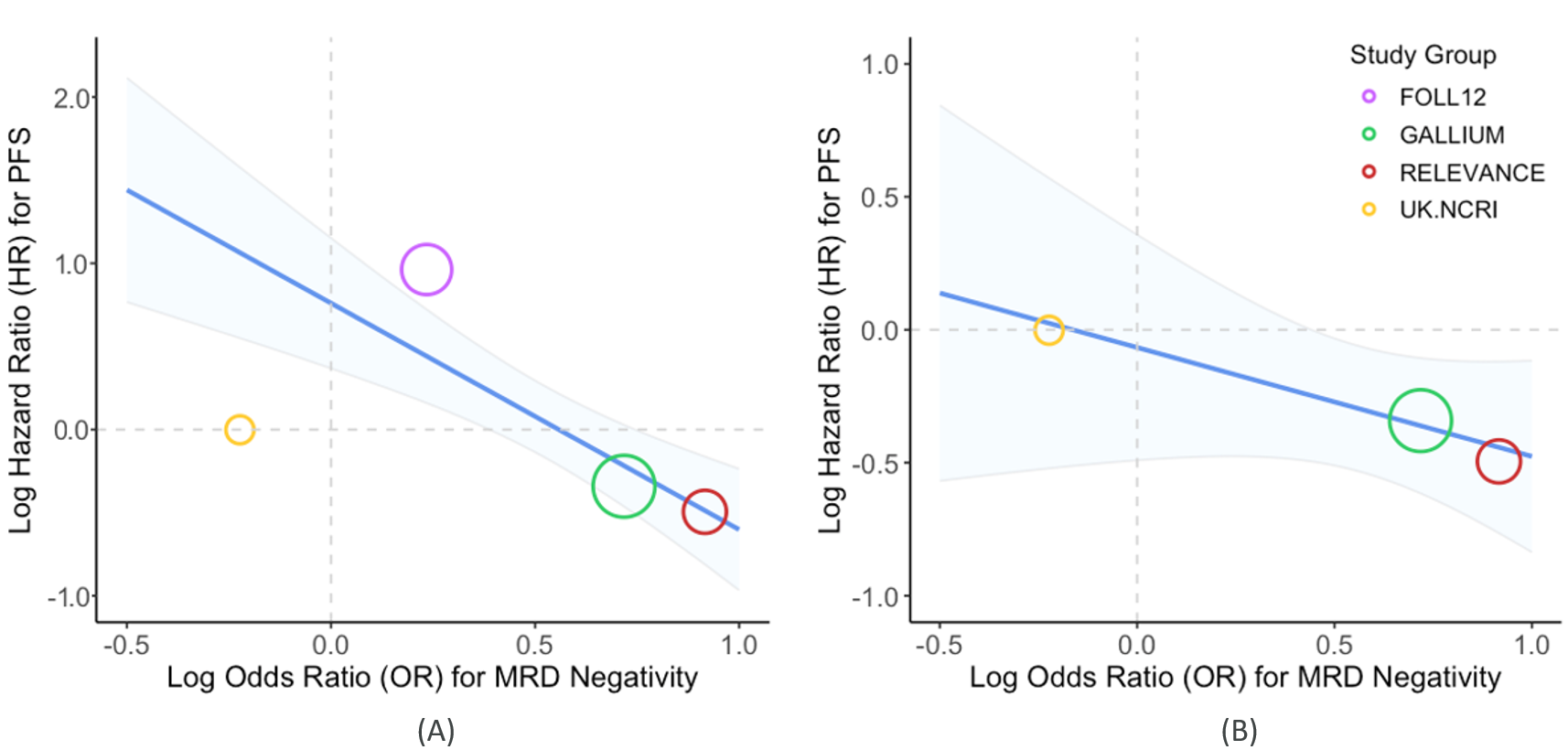} 
    \caption{\textbf{Weighted by sample size}: Correlation between treatment effect on MRD negativity and treatment effect on PFS. (A) all points: WLS for FL with all points included; (B) sensitivity analysis: WLS for FL with FOLL12 excluded}
    \label{fig:FL_wls_N}
\end{figure}

The results in Figure \ref{fig:FL_wls_var} and \ref{fig:FL_wls_N} demonstrate the large impact the FOLL12 trial has on the $R^2_{WLS}$ values, with a significant amount of variation before and after its exclusion. This is more clearly seen in Table \ref{tab:results_FL}, which show wide confidence intervals for $R^2$ estimates, except in the case of the $R^2_{Copula}$ when FOLL12 is excluded. This may be explained by the large heterogeneity introduced by the trial. Since the WLS CIs for $R^2$ are being calculated via bootstrap, this can also explain the wide confidence intervals for those estimates, as the overall number of points is small.

At the individual-level, the global odds ratios can be seen in Table \ref{tab:results_FL}, with significant odds ratios for both the primary analysis as well as the sensitivity analysis excluding the FOLL12 trial. 

Additionally, the $R^2_{Copula}$ gives better precision with a narrower confidence interval, which may be the case due to the Plackett Copula allowing for non-linear relationships between the treatment effect on the surrogate endpoint, and on the true endpoint. Even so, the confidence intervals still remain wide.

\vspace{5mm}
\underline{Acute Lymphoblastic Leukemia (ALL)} \\

The following two-arm papers were evaluated focusing on adult patients: \cite{chen2023allo, zhou2022clinical, yilmaz2020early, jiang2019anti, fracchiolla2023blinatumomab} for B-cell ALL (B-ALL). However, only one of these presented survival curves stratified by both MRD status as well as treatment. Some of these presented only stratification by one MRD status, but not the other, other papers lacked treatment stratification in survival curves. Outside of these papers, there were several publications which had survival curve stratification based on biomarkers \cite{xu2024stat5, boissel2023real, esteve2021allogeneic, cao2023csrp2}, however this was not of interest in our study as we were interested in including treatment effects in our modeling. Without the survival curve stratification by both MRD positive and negative as well as the two treatments, the individual patient data cannot be recreated as only one MRD status or one treatment arm would be present in the data. Therefore, with our current criteria for use, we were not able to extract enough data points for modeling MRD surrogacy in ALL. It is worth noting that \cite{zhou2022clinical, yilmaz2020early} have concluded that MRD status a prognostic factor for survival, so we look to reexamine surrogacy in the adult ALL setting in the future with the availability of data.

As there were not enough two-arm papers, we considered analysis with published one-arm studies, with a summary of papers considered presented in Table \ref{tab:ALL_onearm_lit}. Individual analysis with only OS was considered due to the limited number of studies for the other survival endpoints using data presented in \cite{esteve2021allogeneic, kaito2020allogeneic, zhang2020efficacy, hay2019factors}.

\vspace{5mm}

\begin{table}[h]
    \centering
        \caption{Table of individual-level assessment results for B-ALL}
     \scalebox{0.8}{
    \begin{tabular}{lccc}
        \hline
        Survival Metric & Number of Studies & Settings & MRD Minimum Sensitivity \\
        \hline\hline
        Overall Survival (OS) & 4 & R/R & $10^{-4}$ \\
        Event-Free Survival (EFS) &  2 & R/R & $10^{-4}$\\
        Disease-Free Survival (DFS) &  1 & R/R & $10^{-5}$\\
        Leukemia-Free Survival (LFS) &  2 & R/R & $10^{-4}$\\
        \hline
    \end{tabular}
    }
    \label{tab:ALL_onearm_lit}
\end{table}

\vspace{5mm}

However, some figures did not report overall (non-stratified) survival curves for patients who had MRD measurements, with no information for at-risk patients at each time point. Thus, the OS probability could not be estimated from either a reported survival curve, or through a reconstructed patient curve. 
Therefore, due to the limitations of an even smaller sample size than shown in Table \ref{tab:ALL_onearm_lit}, further analysis was not considered. 

\vspace{5mm}

\underline{Mantle Cell Lymphoma (MCL)} \\

The following two-arm papers were evaluated in our literature search focusing on adult patients: \cite{smith2019minimal, kaplan2020bortezomib,lakhotia2018circulating, pott2010molecular} for MCL. However, none of these manuscripts presented survival curve figures stratified by both MRD status and treatment. Therefore, similar to ALL, we were not able to extract sufficient data points for use in modeling MRD surrogacy in the MCL indication. 

Therefore, we considered one-arm studies, with a summary of papers presented in Table \ref{tab:MCL_onearm_lit}. Both OS and PFS were considered in the 1L setting, as other indications did not have enough trials, with OS considering data presented in \cite{gressin2019phase, tisi2023long, kolstad2014nordic, torka2022ofatumumab} and PFS considering data presented in \cite{gressin2019phase, epstein2024immunochemotherapy, tisi2023long, kolstad2014nordic, torka2022ofatumumab}.

\vspace{5mm}

\begin{table}[h]
    \centering
        \caption{Table of individual-level assessment results for MCL}
    \scalebox{0.8}{
    \begin{tabular}{lccc}
        \hline
        Survival Metric & Number of Studies & Settings & MRD Minimum Sensitivity \\
        \hline\hline
        Overall Survival (OS) & 4 & 1L & $2^{-5}$\\
       Progression-Free Survival (PFS) &  5 & 1L & $2^{-5}$\\
        \hline
    \end{tabular}
    }
    \label{tab:MCL_onearm_lit}
\end{table}

\vspace{5mm}

Within MCL single-arm studies, we encounter the same issues as in ALL, with some manuscripts not including non-stratified survival curves for patients who had MRD measurements, who also lacked information for at-risk patients, rendering us unable to accurately estimate survival probabilities. An additional limitation is the time aspect, where MRD measurements needed to be reported at similar timeframes to allow for comparison of OS/PFS probability at that specific time point. Therefore, we were unable to proceed with single-arm comparisons due to the limited sample size. 

\vspace{5mm}

\underline{Chronic Lymphocytic Leukemia (CLL)} \\

After literature search of $41$ papers involving both MRD measurement and survival endpoints, four datapoints were extracted from the following papers: \cite{dimier2018model, wang2021measurable}. All four trials were phase III and were conducted for patients in the 1L setting. The \cite{dimier2018model} included results from three different trials: CLL8, CLL10, and CLL11 which were extracted.

All trials had a minimum sensitivity of $10^{-4}$ and were measured using flow cytometry. MRD status was measured at the end of induction (EOI) for three studies, and $12$ months post randomization for the E1912 study. The MRD measurement times were reported in Table \ref{tab:char_CLL}. The median follow-up time across all patients is $35.5$ months.

\vspace{5mm}

\begin{figure}[h]
    \centering
        \includegraphics[width=5.9in]{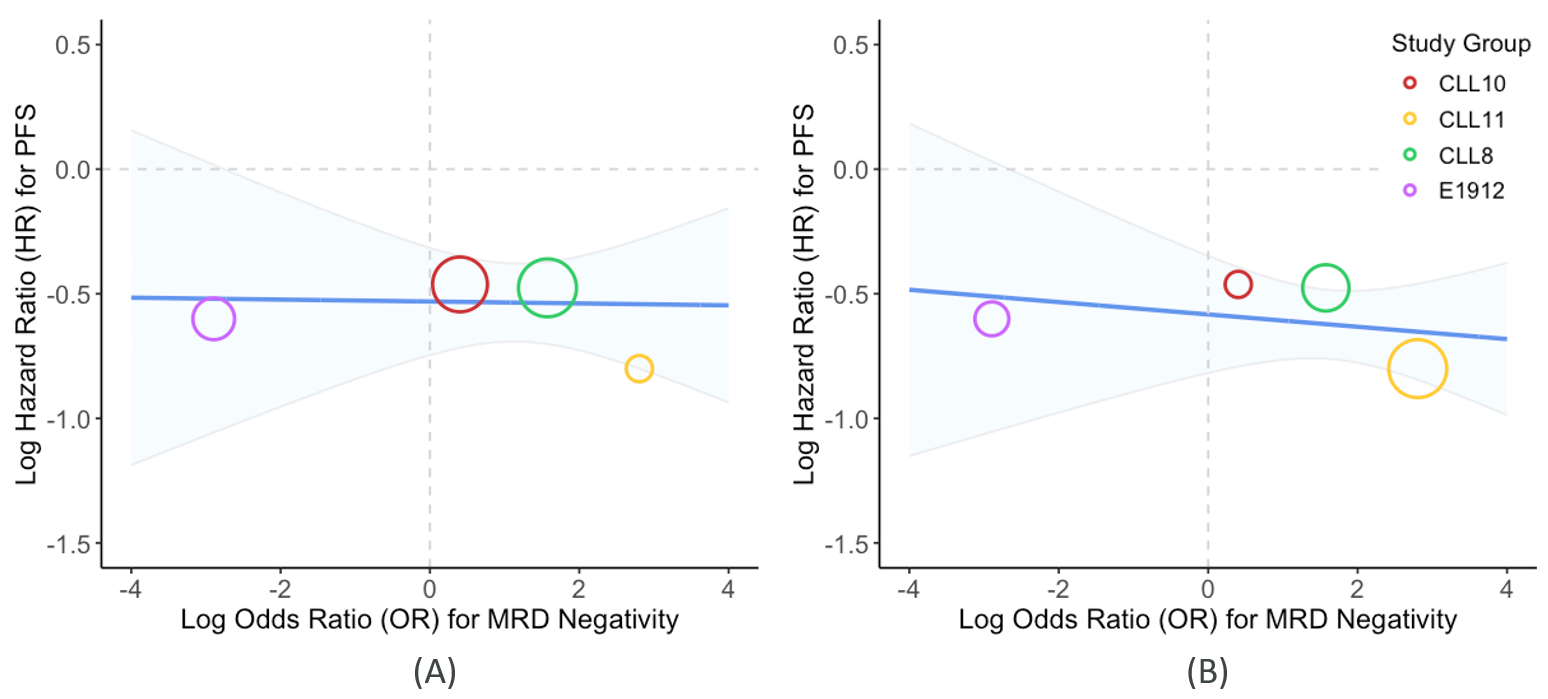} 
         \caption{Correlation between treatment effect on MRD negativity and treatment effect on PFS; (A) weighted by the inverse variance of the log OR; (B) weighted by the sample size}
    \label{fig:CLL_wls}
\end{figure}

\vspace{5mm}

The results in Figure \ref{fig:CLL_wls} show moderate point estimates of $R^2_{WLS}$, however, in Table \ref{tab:results_CLL}, these $R^2$ estimates are shown to have wide confidence intervals in all models. In addition, the $R^2_{Copula}$ is low. As in the case of FL, the wide confidence intervals may be explained by the small number of trials in the analysis.

\begin{table}[h]
    \centering
    \caption{CLL Trial Characteristics and Assessment Results for CLL}
    
    \begin{subtable}{\linewidth}
        \centering
        \scalebox{0.65}{
        \begin{tabular}{lccccc}
            \hline
            Trial Name & Setting/Phase & Sample Size & \multicolumn{1}{p{7cm}}{\centering MRD Measurement Time (months) } &\multicolumn{1}{p{5cm}}{\centering MRD Measurement \\ Method/Sensitivity } & Treatments \\
            \hline\hline
            CLL8  &1L Phase 3 &  392 & 75-195 days after last treatment (EOI)&flow cytometry $10^{-4}$ & FCR vs FC \\
            CLL10 &1L Phase 3 &  337 & 75-195 days after last treatment (EOI)&flow cytometry $10^{-4}$ & FCR vs BR \\
            CLL11 &1L Phase 3 &  474 & 56-190 days after last treatment (EOI)&flow cytometry $10^{-4}$ & G-Clb vs R-Clb  \\
            E1912 &1L Phase 3 &  345 & 12 months after randomization&flow cytometry $10^{-4}$ & Ibrutinib + rituximab vs FCR \\
            \hline
        \end{tabular}
        }
        \caption{CLL trial characteristics}
        \label{tab:char_CLL}
    \end{subtable}

    \vspace{0.5cm} 

    \begin{subtable}{\linewidth}
        \centering
        \scalebox{0.8}{
        \begin{tabular}{lc}
            \hline
            Correlation Metrics & All Trials \\
            \hline\hline
            \textbf{Trial-Level} & \\
            Sample Size & 1548 \\
            $R^2_{WLS}$ inverse variance ($95\%$ CI) & 0 [0, 1.000]\\
            $R^2_{WLS}$ sample size ($95\%$ CI) & 0 [0, 1.000]\\
            $R^2_{Copula}$ ($95\%$ CI) & 0.069 [0, 0.719]\\
            \hline\textbf{Individual-Level} & \\
            Sample Size & 1548  \\
            Bivariate Plackett Copula Global OR ($95\%$ CI) &  10.220 [7.485, 12.954]\\
            \hline
        \end{tabular}
        }
        \caption{Trial/individual-level assessment results for CLL}
        \label{tab:results_CLL}
    \end{subtable}
\end{table}

The individual global odds ratio is shown in table \ref{tab:results_CLL}, with a strong point estimate showing MRD negativity's prognostic effects on PFS. The global odds ratio is significant, as the confidence interval excludes 1.

\vspace{5mm}

\section{Discussion}
\label{sec:disc}

\underline{Follicular Lymphoma (FL)} \\ 

Based on the FL results, before and after sensitivity analysis, it is clear that the conclusions for surrogacy status are highly impacted by the small sample size of the number of trials extracted from the literature, which are fewer than the number used in typical surrogacy analyses. In the primary analysis, the trial-level correlations are weak, and do not exceed $R^2$ values of $0.25$ at the highest. In addition all $R^2$ value have wide confidence intervals, placing a large amount of uncertainty around our point estimates. However, in the sensitivity analysis when the FOLL12 trial is excluded, all $R^2$ point estimates exceed $0.97$, with wide confidence intervals for the $R^2_{WLS}$ estimates. 

As previously mentioned, though there are no FDA guided thresholds for ``acceptable" trial-level correlations, based on our pre-defined criteria, the primary analysis shows weak trial-level correlation. Despite the point estimates of $R^2_{Copula}, R^2_{WLS}$ being high in the sensitivity analysis, the lower bounds of the $R^2_{WLS}$ estimates dip below $0.6$. Therefore, the current data does not meet the criteria of strong trial-level surrogacy. While the sensitivity analysis shows high point estimate for $R^2$ values, the confidence intervals for the WLS estimates are wide, and the sample size is small, casting uncertainty on results based on the current data.

At the individual-level, the global odds ratios for both the primary and sensitivity analyses are statistically significant, as the $95\%$ CIs exclude $1$. This can essentially have the interpretation that achievement of MRD negativity is prognostic for PFS. However, based on our pre-specified criteria of association strength, we can only consider the resulting association as ``moderate", as the global ORs reach only $2.7$ at the highest, which is below the threshold of $3$ for defining a ``strong" association.

Therefore, there is no current evidence that MRD-based endpoints can be used in place of PFS in FL. This study is limited by the small sample size in terms of number of trials included, so the surrogate threshold effect was not calculated. \\

\underline{Acute Lymphoblastic Leukemia (ALL)} \\ 

The B-ALL data was not able to be extracted, so we cannot make any conclusions about MRD surrogacy in this particular indication. However, \cite{galimberti2018validation} studied validation of MRD as a surrogate for EFS in pediatric ALL and concluded that there was no evidence both at the trial-level and individual-level that MRD could be a surrogate endpoint for EFS. There is a need to examine MRD surrogacy in the adult ALL setting, however, with the currently mentioned limitations of existing literature, it is difficult to perform such an assessment. \\

\underline{Mantle Cell Lymphoma (MCL)} \\ 

The MCL literature search also yielded no extractable points for our analysis. Due to the previously mentioned challenges in using single-arm studies, single-arm analyses were not able to be incorporated. To our knowledge, there have not been existing analyses of surrogacy in the MCL indication. This would be a valuable future field of exploration. \\

\underline{Chronic Lymphocytic Leukemia (CLL)} \\ 

The trial-level correlations were weak in CLL, with an $R^2_{Copula}$ value of only 0.069. In addition, all lower bounds for the $95\%$ CIs were smaller than the prespecified threshold of $0.6$. Therefore, the current data does not meet the trial-level surrogacy criteria. 

However, the individual-level, the global odds ratio is strong, exceeding the specified criteria of an OR of $3$. The $95\%$ CI excludes 1, so the odds ratio is statistically significant.

While MRD is therefore not a fully validated endpoint, there may be useful information due to the strong individual-level association. While there is almost no correlation between treatment effects on MRD and treatment effects on the PFS, MRD can be interpreted as highly prognostic for PFS. Meaning, patients who are MRD negative have longer survival outcomes. But, that MRD may not be predictive of PFS, as it fails to capture treatment effects on the MRD-based endpoint. However, due to the small number of data points, the authors would like to exercise caution in the interpretation of the overall analysis. Due to the small number of data points, the surrogate threshold effect was not calculated.

\section{Conclusion}
\label{sec:conc}

For both FL and CLL, there was no evidence for MRD to be considered as a fully validated endpoint due to the absence of a strong trial-level correlation. However, in both indications, individual-level correlations were significant, meaning, MRD can be prognostic for PFS in both FL and CLL, but may not be predictive. 

In general, one of the major limitations is that the sample size in terms of the number of usable trials was small in all indications which means that our estimates have wide confidence intervals and may not be robust, or we may not have been able to perform analysis. Additionally, due to the small number of trials, trials with larger sample sizes primarily drive inference, as they carry more weight. We would encourage a revisit of this problem when more literature is available.

An additional limitation of meta-analytic studies, and this study in particular, is that data in manuscripts must be presented a certain way for extraction. Particularly, survival curves must be stratified by both positive and negative MRD status as well as by the treatment in order to extract the individual patient data. In one arm trials, in order to estimate a level of correlation between the true endpoint and the surrogate, also need survival curves stratified by MRD status, as well as reporting the number of patients at risk at each time point. Without this information, censoring considerations are neglected, if only the total number of patients per arm at the measurement time are presented.

There may be patient heterogeneity due to differences in disease stage, genetic mutations, and other characteristics. MRD timing varies as well, as not all studies in an indication measured MRD, for example, at EOI. In addition, different studies use different assays to measure MRD, and may also have differing levels of sensitivity which can be seen in the tables for each of the indications above.

We also encourage looking through the lens of multivariate meta-analyses which may broaden the scope of literature search by including studies that do not report the main outcome of interest, but a correlated outcome \cite{bujkiewicz2019nice}.

This meta-analysis provides valuable insight into the current scope of literature in the hematology oncology trials regarding reporting on the use of MRD. The authorship team hopes that the limitations of the current reporting characteristics encourage further exploration of MRD as an endpoint and sharing result of MRD use in order to promote a robust look at potential MRD surrogacy down the line.

\section{Funding Statement}
\label{sec:ack}

The authors would like to thank Oncology Biometrics at AstraZeneca PLC for the research funding. Thank the IP team at AstraZeneca PLC, specifically Sushma Kavikondala, Nicola Rath, and their team for their support in conducting literature reviews and categorization. Thank Binbing Yu and Dan Jackson from the statistical innovation group for their consultation on the analysis methods.

\section{Conflict of Interest}

None to report.

\section{Author Contributions}
J.S., X.C., M.I., and J.L. contributed to the design and implementation of the research, to the analysis of the results, and to the writing of the manuscript.

\section{Supplemental Material}

\underline{Details About Surrogate Endpoints}

Surrogate endpoints have been defined by the National Institutes of Health (NIH) as biomarkers that are intended to substitute for a clinically meaningful endpoint that are expected to predict the therapeutic effect and can be measured earlier \cite{united1992new}]. Surrogacy was formalized in Prentice’s seminal paper \cite{prentice1989surrogate}, which discuss formal statistical criteria for surrogacy. However, these
criteria were not simple to verify, and have since been expanded upon since introduction \cite{zhao2016surrogate}.
Currently, one of the more common methods for evaluation of surrogacy is guided by Buyse et al’s recommendation to study the individual-level associations between endpoints and the trial-level associations, with surrogacy being validated when strong associations are demonstrated in both \cite{buyse2000validation}. To examine individual-level association, the correlation between the surrogate endpoint and the true endpoint can be estimated using a bivariate copula model, as described in \cite{burzykowski2001validation, burzykowski2004validation}, which accounts for trial-specific treatment effects as well as individual-level correlation. The resulting value of interest is a ratio of odds ratios, known as a “global odds ratio” accompanied by a confidence interval. The global odds ratio is the ratio of (a) odds of experiencing an event (viz. PFS) at or beyond a certain timepoint for patients who are MRD negative and (b) the odds of experiencing the same event at or beyond the same timepoint for patients who are not MRD negative.

Trial-level associations can be modeled by relating the magnitude of the treatment effects on the true and surrogate endpoints through a linear regression model which may be weighted by sample size \cite{buyse1998criteria} as well as the previously discussed bivariate
copula model, with the trial-level correlation measured in terms of an $R^2$ value.

Additionally, at the individual-level, a global odds ratio is considered statistically significant if the $95\%$ CI excludes 1, with a larger magnitude of the point estimate corresponding to a stronger association, and we consider point estimates of the global OR $\geq$ 3 as a strong association.
In addition to the individual-level and trial-level surrogacy measures, a surrogate threshold effect (STE) can be calculated, which is the required minimum treatment effect on the surrogate endpoint required to predict a non-zero treatment effect on the true endpoint in future trials \cite{burzykowski2006surrogate}. This is typically done after the surrogate endpoint is quantified based on the individual and trial-level assessments   to have potential for surrogacy. The STE is computed from a linear regression model between the treatment effects using the point where the lower limit of $95\%$ prediction interval intersects the no-treatment effect line. This gives an idea of what the minimum effect would be needed in future trials.

While surrogacy has the potential to speed up the timeline for patients to access new medications, it also runs the risk of misleading the benefits of therapy. In the case of bevacizumab, which was granted accelerated approval by the FDA for first-line treatment for HER-2 netative metastiatic breast cancer, the surrogate endpoint was PFS, with the true endpoint being OS. However, later confirmatory trials failed to demonstrate benefits in the true endpoint, OS, so the drug was withdrawn from the market \cite{carpenter2011reputation}. Therefore, validation of surrogacy must be performed carefully without overreaching the conclusions.

In this analysis, a fixed-effects model was chosen which uses additive rather than multiplicative model error assumptions due to the small number of trials \cite{tufanaru2015fixed}. Fixed effects modeling was selected over equal-effects modeling, as equal-effects modeling assumes that true outcomes are homogenous, while the fixed-effects model does not make this assumption and instead models a weighted mean of the true outcomes \cite{viechtbauer2010conducting}.

\underline{Data Extraction Validation}

Hazard ratios reported in the paper were re-calculated as a form of quality control, with estimation of the parameter of interest deemed acceptable if the values were close. For example, if a paper only reported the hazard ratio of PFS in one treatment arm, this hazard ratio was recalculated, despite the value of interest being the hazard ratio of one treatment arm as compared to another; if this recalculation was similar, we can have confidence in the accuracy in the estimated value of interest. The ``IPDfromKM" software itself has a few measures of validation such as the root mean squared error (RMSE), mean absolute error (MAE), and a Kolmogorov-Smirnov (KS) test procedure. The KS test procedure tests the null hypothesis that the distributions of the input and reconstructed (estimated) survival probabilities are the same and outputs a p-value; however, the limitation of this procedure is that a large sample size results in almost automatic test significance when there are even trivial deviations from the distribution. Thus, the need for validation through reconstructing other hazard ratios reported in the paper as well as visual reconstructions of the survival plots.

\underline{One-Arm Studies}

\begin{figure}[H]
    \centering
    \includegraphics[width=\linewidth]{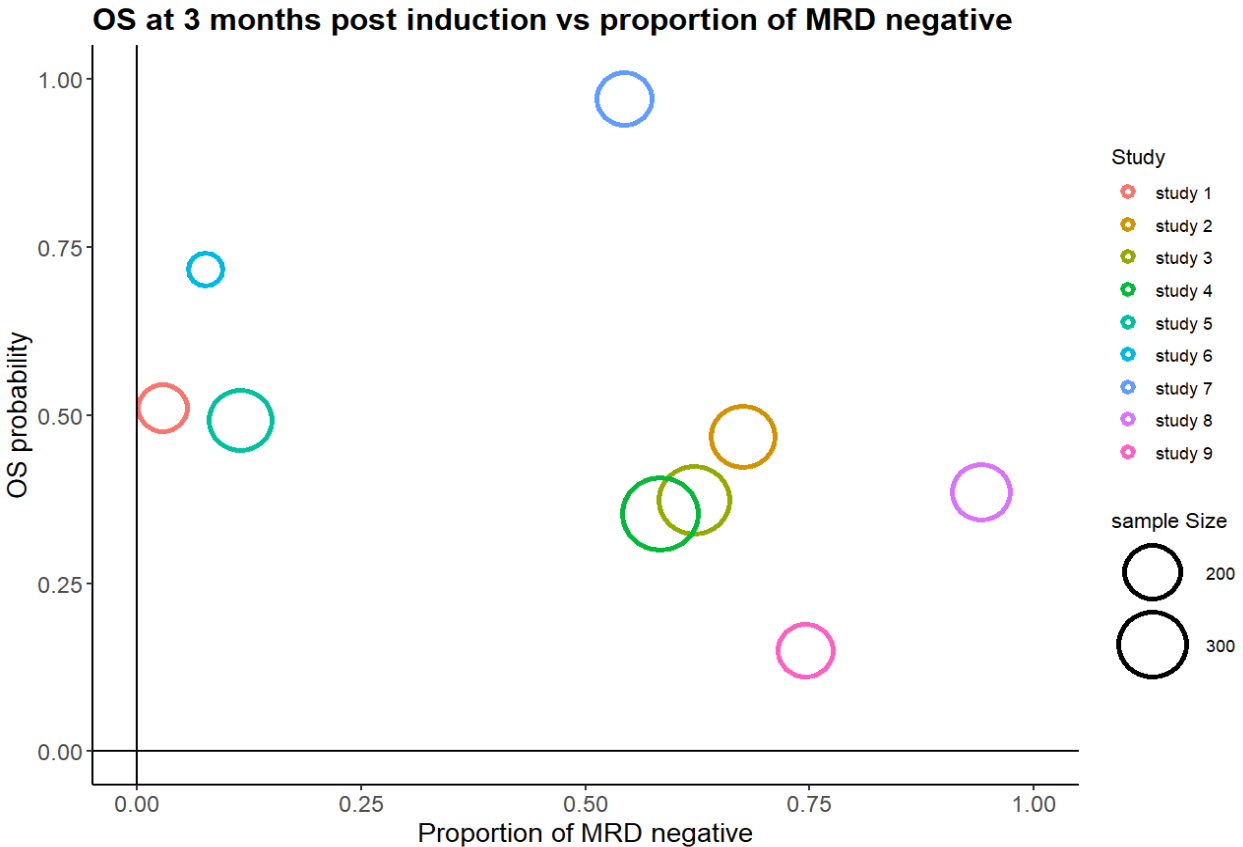}
    \caption{Example Plot of Single Arm Correlation}
    \label{fig:sing_arm}
\end{figure}

For the indications which lacked two-arm trials (MCL, ALL), we considered use of one-arm trials. While this proposed method cannot be used to make any conclusions about surrogacy due to the absence of a treatment effect, the idea is to estimate an association between PFS/OS at a certain time point and the probability of MRD negativity, which may show the correlation between the true endpoint (PFS/OS) and MRD negativity. An example can be seen in Figure \ref{fig:sing_arm}. However, readers should be cautious about the interpretation, as such a correlation may not reflect treatment effect. In addition, this proposed method may be difficult to employ, as the survival probabilities need to reflect survival probabilities at the same time point (ex. 3 months after EOI) in order to have an accurate comparison; it may be difficult to compile such a list of datapoints, unless the available literature is relatively vast, and MRD measurements are relatively standardized in their time frames. Techniques for proportion-based meta-analysis have been previously utilized, such as in \cite{lin2020meta, unknown}, and have had packages written in R. Bayesian methods exist as well \cite{singh2021incorporating}, however, we do not take this approach due to the small sample size. 

\vspace{5mm}

\newpage

\cite{wang2017validation} employed meta-analytic surrogacy evaluation in one-arm trials by creating artificial treatment groups using risk factors. Patients were stratified by the risk factor into a high risk and low risk group, effectively mimicking the treatment effect for PFS surrogacy evaluation, as the authors hypothesized that the risk factors were prognostic in mesothelioma. However, such an approach was not possible in this analysis, as there are no methods for extracting covariates for patients from survival curves-- only their survival times, failure time indicators, and MRD status, if provided as an existing stratification. 



\bibliographystyle{apalike}
\bibliography{refs}

\end{document}